# Polyp Segmentation in Colonoscopy Images using U-Net-MobileNetV2

Marcus Venicius Lau Branch[1], Adriele dos Santos Carvalho[2]
[1]Federal University of Santa Catarina, Brazil
[2]Federal University of Santa Catarina, Brazil
engmarcusbranch@gmail.com

*Abstract*—Colorectal cancer from the appearance of polyps that can be benign or malignant is one of the most fatal diseases in the world. To find these polyps in patients, colonoscopy is performed, which is a very efficient technique in this case. Clinically, detecting and segmenting these polyps in order to determine their presence or not is a very difficult process that requires a lot of time and experience from professionals, depending directly on these factors. Therefore, it becomes increasingly important to have an automatic, effective and reliable method of detecting and segmenting these polyps, making diagnoses faster and more accurate. In order to assist in the development of a method, we proposed the U-Net-MobileNetV2 model, which is the combination of two neural networks, where one acts as an encoder for the other and is responsible for learning image resources. Our experiments generated satisfactory results, demonstrating a good performance and good segmentation. U-Net-MobileNetV2 achieved a Dice Coefficient of 89.71% and an IoU of 81.64% for the Kvasir-SEG dataset, where both are higher than the results obtained by other state-of-the-art models.

## 1 INTRODUCTION

According to PAHO (Pan American Health Organization) colorectal cancer has a significant rate of deaths related to this disease worldwide. The presence of this type of cancer is assessed when there are polyps (a projection of tissue growth from the wall of an empty space, such as the intestine, which can be benign or malignant, and can even develop into cancer) therefore, it is extremely important that these polyps are discovered in the early stages by doctors and this investigation occurs through the colonoscopy exam. According to [1] many polyps are lost or not observed during the exams, as it depends a lot on the experience of the doctor who performs it, so a method that is able to detect and segment these polyps will automatically help in the diagnosis thus decreasing the chance that some polyps will be forgotten and, consequently, wrong diagnoses will be made.

Colonoscopy is an invasive exam that captures real-time images of the large intestine and part of the terminal ileum (the final portion of the small intestine). To perform the exam, a device called a colonoscope is used, which has a thin and flexible tube with a camera at its end capable of filming the inside of the intestine in order to investigate the presence of colorectal cancer, polyps and inflammatory bowel diseases.

The main objective of this research is to perform the segmentation of polyps present in frames extracted from colonoscopy videos, which have several examples of these anomalies. To solve this problem, the U-Net [2] architecture will be used with the MobileNetV2 [3] network, with individual tests being carried out for each network.

This research is divided as follows, in the section 1 the introduction and the problems that started this research are presented, while the section 2 brings other works related to the segmentation area of biomedical images, for the section 3, the approach adopted to perform the polyps segmentation is reported, as well as the presentation of the network model implemented, then in the section 4, the results of the implementation of the networks on the dataset are exposed, then in the section 5, the results obtained by the model are analyzed and discussed and at the end, in section 6, the conclusions of this research are presented and also ideas for future work.

## 2 RELATED WORKS

The application of Machine Learning and Deep Learning techniques in the segmentation of biomedical images is something that is in great evidence and growth today. There are different types and formats of medical images that are used, among the most used in the state of the art are magnetic resonance, x-ray and ultrasound images of different organs of the body. As mentioned in the previous section, this project will use images taken from colonoscopy videos, that is, exam frames in which the purpose is to perform the segmentation of polyps through a supervised learning approach.

The approach to perform the segmentation differs according to the results you want to obtain. The segmentation can work as a binary classification, in which a certain part of the image will be segmented, probably receiving the label 1 and the rest will be the bottom of the image, receiving the label 0, it can also be multi-class, when the target in the image is multiple or a specified group, etc. As stated earlier, in this research we will use supervised learning, as the data set has images to be segmented and the corresponding masks.

The segmentation of an image is the process of classifying each pixel as belonging to a previously defined label. In the last few decades, there has been a lot of research related to the segmentation of different images with different content. The first methods implemented for segmentation of polyps such as [4] and [5], trained classifiers to distinguish one polyp from the rest of the image, however these models have a high error rate. Currently, the vast majority of models used to perform segmentation are convolutional neural networks or FCNs (Fully Convolutional Networks) with a pre-trained model to identify and segment polyps [6][7] and this is explained by the significant improvement obtained when these models are implemented. A more current approach present in the state of the art is the one presented in [8], in which the authors used a deep neural network for segmentation, valuing the area and limit of the polyps to identify these anomalies, since they have different sizes and formats, then analyzing these two characteristics helps a lot in the hit rate.

With the success of U-Net, proposed by [2], when applied to the segmentation of medical images, [9] applied this architecture with some modifications and focused on the segmentation of the entire area polyp and obtained promising results, however they ignored the restriction of the area limit or size of the anomaly, which may end up depreciating the model's performance in segmentation.

From the original U-Net architecture that consists of encoder-decoder, it is known that the encoder extracts characteristics (or resources) from the input images and these resources are concatenated with the decoder so that the segmentation task is performed by the network. Using pre-trained networks in large databases in these cases helps a lot when performing segmentation since your weights will already be trained with millions of learned resources and the weights will not start from scratch during training.

## 3 APPROACH

### 3.1 Dataset
The dataset used was the Kvasir-SEG [10], which has 1.000 images of polyps and their corresponding masks segmented by experienced doctors.

In the training conducted during this research, the data will be separated into training, validation and test sets with a proportion of 80%, 10%, 10%, respectively.

### 3.2 Pre-processing
For the images of the data set to be used, two pre-processing steps were applied. The first was the normalization of the pixels of the images in which these pixels that are in the range of 0 to 255 are reset to a range of 0 to 1, avoiding slowness in training and learning, since the data was in its raw format. The second was the implementation of data augmentation techniques such as CenterCrop, RandomRotate, HorizontalFlip, VerticalFlip and GridDistortion.

### 3.3 U-Net-MobileNetV2
Figure 1 shows an overview of the proposed U-Net-MobileNetV2 network where a pre-trained MobileNetV2 with the weights of the ImageNet database [11] is applied as the encoder. The proposed model receives the input images with a size of 320x320, which are inserted in the pre-trained encoder, which is based on inverted residual blocks (or structures), which includes a combination of spatial convolutions with 3x3 kernels, ReLu activation and layers of Batch Normalization. In this case, the encoder will use compact depth-to-depth convolution to filter and learn characteristics of images that feed the network. The use of these inverted residual blocks helps to reduce the number of parameters, making the model easier and faster to train. Another advantage is that the model will perform better and converge faster than if a pre-trained network was not being used. In the decoding path, up-sampling operations are used to increase the size of the feature map back to its original size. During this path, the characteristics are concatenated between the encoder and decoder blocks and also go through a 3x3 convolution layer, followed by Batch Normalization and ReLu activation. Finally, the last block of the network is a block with a 1x1 convolutional layer and a Sigmoid activation, so that the segmentation performed by the network can be generated.

### 3.4 Implementation details
The proposed model was implemented with the Keras [12] framework and TensorFlow [13] as a backend. The data set was divided into subsets of training, validation and testing with a proportion of 80%, 10% and 10%, respectively. The trainings were carried out with a learning rate of 0.0001, batch size of 16, since a lower or very small value could lead to overfitting, the chosen optimizer was Adadelta and the model was trained for 100 epochs and converged with 76 epochs. Early-stoping regularization was also implemented in the validation subset, specifically in the validation loss as another way to avoid overfitting.

The metrics chosen to assess the model's performance were the Dice Coefficient (DSC), and the Intersection over Union (IoU). As a loss function, Dice Loss was implemented, as it generates better results in segmentation tasks, even though some cases in the state of the art still carry out experiments with loss functions such as binary cross entropy. A threshold value of 0.5 was used to convert the pixels to background or foreground. The proposed network was implemented using a single XFX Radeon RX 580 GTS GPU with 8 GB of memory. The complete training of the network took 4 hours.

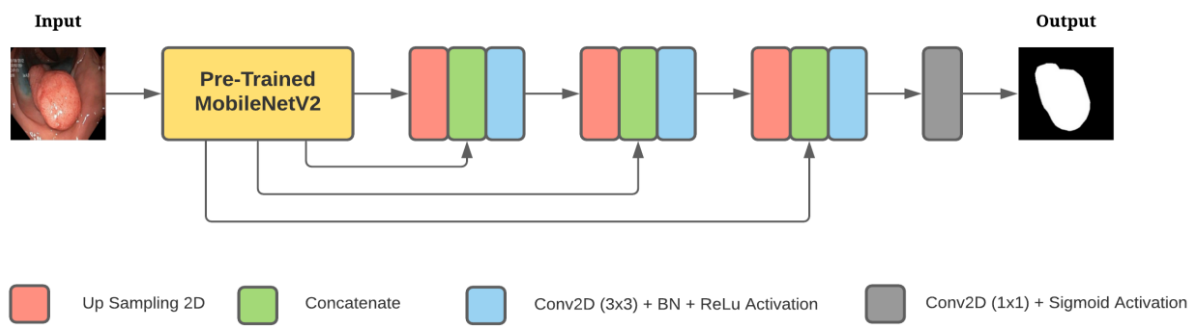

**Fig. 1:** Proposed U-Net-MobileNetV2 architecture.

## 4 RESULTS

To evaluate the performance of the model, the experiment with the Kvasir-SEG dataset was conducted. For comparison of results, we compared the results obtained by the proposed model with the results of other three state-of-the-art models that were also proposed for this polyp segmentation task. For evaluation and comparison between the models, the values of the Dice Coefficient and the IoU were exposed. Table 1 presents the results of the models U-Net-MobileNetV2, ResUNet [10], ResUNet++ [14] and an unnamed model proposed in [15] on Kvasir-SEG.

**Table 1:** Quantitative results of all models on Kvasir-SEG.

| Method | Dice | IoU |
| --- | --- | --- |
| **U-Net-MobileNetV2** | **0.8971** | **0.8164** |
| Tomar et. al [15] | 0.8411 | 0.7565 |
| ResUNet++ [14] | 0.8133 | 0.7927 |
| ResUNet [10] | 0.7877 | 0.7777 |

Table 1 shows that the U-Net-MobileNetV2 model obtained the highest Dice Coefficient and also the highest IoU among the models analyzed for the Kvasir-SEG dataset. Based on the models proposed in [10], [14] and [15], in the Kvasir-SEG dataset, U-Net-MobileNetV2 achieved better results with percentage averages of 13.888%, 10.303% and 6.657% when analyzing the Dice Coefficient and of 4.976%, 2.989% and 7.918% when analyzing the IoU.

The proposed model surpassed the baseline architecture [10] in terms of all metrics, mainly in terms of the Dice Coefficient, which was surpassed by a large margin. We can say that this margin of difference between the U-Net-MobileNetV2 and the other state-of-the-art models indicates that the use of a pre-trained network as a U-Net encoder is able to optimize the performance of the model in the segmentation task and is able to reduce computational cost and training time. Figure 2 presents some samples of the segmentations obtained by the proposed model, which we refer to as qualitative results. In Figure 2 we have samples of original images, their corresponding masks and beside, the mask predicted by U-Net-MobileNetV2.

## 5 DISCUSSIONS

The proposed U-Net-MobileNetV2 obtained satisfactory results in the Kvasir-SEG dataset (see Table 1). From Figure 1, it is clear that the segmentation generated by the proposed model is better than those generated by the other state of the art models used for comparison, as it was able to better identify the shape and location of the polyps in the images. In general, our proposed method generated a segmentation mask that was closer to the ground truth provided by the dataset than the other models.

During the training of the network, a manual tuning of hyperparameters was performed, so that we could find the best combination of parameters and the best result. With this search for hyperparameters, it was evident the influence that values such as learning rate, batch size, number of epochs and optimizer can have on the model, being able to depreciate or improve performance.

Analyzing the performance of the model, we understand that the model and the results can be improved using other pre-trained networks as an encoder, implementing more data augmentation techniques or using a dataset with a larger number of samples, performing more tests with new hyperparameters and implementing some post-processing techniques so that the segmentation generated is even better and closer to the ground truth.

The proposed model learns the information from depth-to-depth convolutions and filters this information to each block and stage of the network, allowing the creation of more refined resource maps, where the most relevant resources of the images are extracted, which ended up helping to increase the efficiency of the segmentation.

U-Net-MobileNetV2 can and should also be applied to segmentation tasks of other types of images and not only to medical images, since it will probably also perform well even if for other images the pixel classification needs more detailed validations, as well as apply to different class targets and not just binary pixel classification.

## 6 CONCLUSIONS

In this paper, we proposed and presented the U-Net-

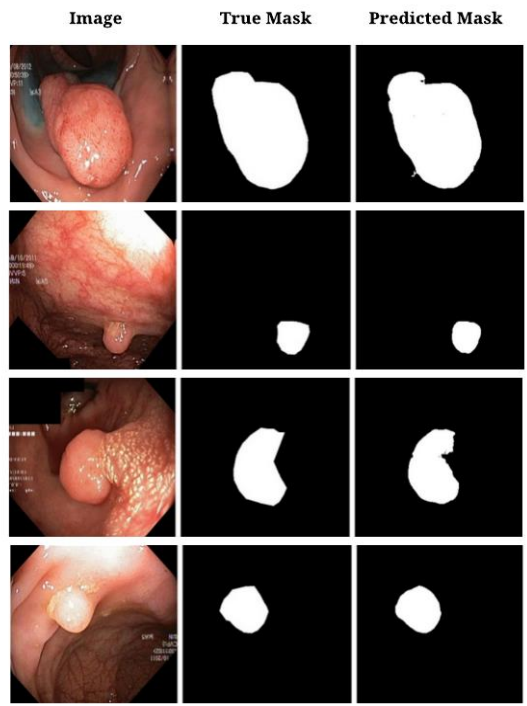

**Fig. 2**: Qualitative results of U-Net-MobileNetV2 on Kvasir-SEG dataset.

MobileNetV2 model for the automatic segmentation of polyps in colonoscopy exam images. Our model and the successive tests that were carried out demonstrate that the proposed approach outperformed the models present in the state of the art that were cited and used for comparison. Very representative and competitive values were achieved for the segmentation task with the Kvasir-SEG dataset as we reached a Dice Coefficient of 0.8971 and an IoU of 0.8164. A great advantage of this model that helped a lot in training was the implementation of inverted residual blocks, which make training faster and more efficient, as well as making the model converge more quickly.

The proposed network can serve as a basis for future investigations and more in-depth studies, with the aim of developing more and more an automatic method of segmentation of polyps that is reliable. From this research, we intend in the future to also test different pre-trained networks as an encoder so that a broader comparison can be made between different models with different networks previously trained in the encoder, as well as to apply post-processing techniques that further improve the segmentations generated by the model and test the model's performance in different datasets with new data so that the generalization of the network can be analyzed. Finally, we hope that this study can provide the community with an additional information base and future research opportunities that further explore segmentation tasks.